# Emotive Architectures: The Role of LLMs in Adjusting Work Environments

L. Vartziotis[1,2] T. Vartziotis[1,2], F. Beutenmüller[2], S. Salta[1], K. Moraitis[1] M. Katsaros[1], Sotirios Kotsopoulos[1,3]

[1]National Technical University of Athens, Greece

[2]TWT GmbH Science & Innovation, Stuttgart, Germany

[3]Massachusetts Institute of Technology, Cambridge USA

**Abstract**

In remote and hybrid work contexts, the integration of physical and digital environments is revolutionizing spatial experiences, collaboration, and interpersonal interactions. This study examines three fundamental spatial conditions: the physical environment, characterized by material and sensory attributes; the virtual environment, influenced by immersive technologies; and their fusion into hybrid environments where digital and physical components interact dynamically. The increasing number of AI tools in contemporary society, extensively utilized in both professional and personal spheres, has led to a varied landscape of developing technologies. For instance, ChatGPT has emerged as one of the most downloaded applications, a statistically substantiated fact that demonstrates the swift incorporation of language-based AI into daily life. It also underscores the function of large language models (LLMs) as meaningful bridges between concepts at reading emotional and behavioral signals via natural language. These models provide real-time modifications such as altering illumination, acoustics, or interface configurations, converting static settings into dynamic, emotionally receptive environments. We investigate the integration of language models into professional settings and their potential to enhance user experience by promoting focus, well-being, and engagement. The study investigates ethical concerns, including privacy, emotional tracking, and user agency, emphasizing the importance of inclusive and transparent design. This research formulates a framework for creating co-adaptive environments that merge technological innovation with human-centered experiences, offering a fresh viewpoint on responsive and supportive hybrid workspaces.

**Introduction**

The fusion of digital and physical environments refers to the integration of physical areas with virtual and computationally enhanced layers, resulting in interactive and adaptive hybrid environments. This integration is manifested in robotics, mobile modules, and dynamically adjustable architecture that react instantaneously to user presence and activity, facilitating responsive workspaces and adaptable learning environments that mirror the attributes of the physical world. In the digital domain, powerful generative AI—such as large language models (LLMs), image synthesis systems like DALL·E, and sophisticated video recognition algorithms—facilitates real-time content generation and contextual understanding. These technologies provide immersive simulations, automated asset generation, and multimodal interaction scenarios, significantly expediting iteration and customization in XR design processes (Guo et al., 2023).

The genuine potential arises from the integration of environments: systems that not only superimpose data but also actively analyze and react to emotional signals, cognitive conditions, and interaction dynamics. Recent studies on VR-based emotion tracking utilizing biosignals and visual inputs reveal the ability to ascertain user states, including valence and arousal, in real time, with enhanced precision achieved by multimodal sensor integration. (Chung et. al., 2023). Projects such as the Magic XRoom platform exemplify controlled emotion elicitation alongside recognition to adapt environment difficulty and ambiance in real time (Chung et. al., 2023). However, studies have shown that while AI can process posture and proximity, it still lags behind humans in interpreting nuanced social interaction without explicit programming (Zhao et al., 2021). By embedding emotion-recognition engines in everyday workspaces—using embedded cameras, voice-analysis tools, and physiological sensors—organizations can monitor metrics like mood, engagement, and stress over time. Such systems could trigger alerts when satisfaction dips or detect patterns such as afternoon fatigue or meeting overload. Experimental frameworks in VR counseling also suggest generative LLM-driven avatars can bolster well-being and promote reflective practices (Kim et al., 2022). Yet, emotion AI raises ethical and legal considerations: real-world deployments face concerns about privacy, accuracy, and fairness, as highlighted by the EU's proposed restrictions on emotion AI in sensitive settings like education and unregulated workplaces (Wachter and Mittelstadt, 2023).

In summary, the convergence of robotic adaptability, generative virtual content, and emotion-tracking systems could redefine remote work, collaborative design, and virtual conferencing. However, realizing this vision demands rigorous technological validation, ethical safeguards, and transparency—ensuring that AI-enhanced environments support not only productivity and creativity but also human dignity and well-being.

**Definition of Workspaces**

This paper examines and defines three key types of environments shaping today's remote and hybrid workspaces: the physical environment, the virtual environment, and their fusion into hybrid spaces. Drawing from recent research, we explore how Large Language Models (LLMs) act as semantic mediators within these environments—providing a statistical interpretation of emotional and behavioral cues to trigger adaptive changes in both physical and digital space. We analyze the impact of this integration on the user experience, showing how LLMs can enhance attention levels, reduce stress, and create more responsive, emotionally engaging work environments.

*Physical environment*

The real environment is a culturally framed and dynamically experienced field where natural and built elements intertwine. It is constructed not just physically, but through symbolic, sensory, and participatory acts that engage both memory and material, form and transformation. Architecture has always been shaped by technological change—from the Stone Age to the twentieth century, where electric lighting and elevators enabled new urban forms. Modernist pioneers like Adolf Loos, Le Corbusier, and Mies van der Rohe embraced industrial materials and technologies, aiming for efficient, rational forms. Their vision of the home as a "machine for living" reflected a broader faith in progress and systematisation (Frazer, 1979). However, this trajectory evolved. By the mid-20th century, architecture began to shift from static, object-based designs toward systemic, dynamic frameworks. British architect Cedric Price exemplified this shift by reimagining architecture as an instrument of system intervention--a means of facilitating human processes, not dictating form (Kousoulas and Perera, 2021). His work, particularly the Inter-Action Centre (1970-77), marked a key moment in architectural thinking where the built environment was seen as a responsive system embedded in the social and physical environment (Herdt, 2021).

Price, influenced by cybernetics and systems theory (notably through collaborations with Gordon Pask), developed a flexible architecture that could adapt to changing needs. His spatial strategies used surveys, modular design, and temporal diagrams to reflect evolving community dynamics (Herdt, 2021). Such projects redefined architecture not as a fixed container, but as part of a larger ecological system—a dynamic interplay of space, technology, and social interaction. Rather than reinforcing static spatial hierarchies, Price's ecological approach positioned architecture as a performative and relational environment, blurring boundaries between design, use, and transformation. His vision continues to inspire how we understand the physical environment not merely as backdrop, but as an active participant in shaping human experience and community life (Kousoulas and Perera, 2021). Contemporary architectural discourse increasingly challenges the classical notion of space as a neutral or static background for human activity. In light of shifting socio-technical paradigms—including remote work, smart infrastructures, and immersive computing—the physical environment is now understood as a relational system rather than a predefined object. This transition fits with rising aspirations for responsive, inclusive, and culturally rooted spatial design capable of embracing technological augmentation.

The physical environment serves as a dynamic intermediary among users, materials, and systems, rather than only acting as a static substrate for digitization or overlay. It is both corporeal and interpretative: molded by its material attributes and the experiential, emotional, and cultural connections it maintains. This idea offers the foundation for a meaningful integration of artificial intelligence or virtual frameworks into spatial experience. Several architectural ideas have contributed to this broadened vision of physical space, each viewing it as an emergent and contingent condition shaped via interaction, variation, and perception. Yona Friedman's notion of mobile architecture envisions the built environment as a participative framework, adaptable to the evolving requirements of its users. In his unrealized theoretical constructs Pro Domo and Ville Spatiale, Friedman advocates for architectural designs that transform via occupancy, highlighting the significance of agency, adaptability, and indeterminacy in form (Friedman, 2006). Bernard Cache (1995) presents a complementary but unique topological technique. In *Earth Moves*, he articulates architecture art of inflection, asserting that design encompasses the modulation of forces, behaviors, and spatial tendencies rather than solely the generation of shapes. For Cache, the physical environment is a territory of continuous variation, shaped by the interplay of conditions rather than bounded by formal constraints.

Additional thinkers reinforce this conception of space as sensory and situated. Juhani Pallasmaa (2005) foregrounds embodied perception, advocating for architecture that engages all senses and supports psychological well-being. Peter Zumthor (2006) articulates the atmospheric character of space—its capacity to evoke memory, emotion, and presence through material, light, and proportion. Alberto Pérez-Gómez (1983) situates architecture within a symbolic and poetic lineage, contending that spatial comprehension is invariably influenced by cultural narratives, metaphors, and imaginative constructs. Kenneth Frampton's theory of critical regionalism emphasizes local context, material specificity, and cultural rootedness, creating a framework for resisting technological homogeneity (Pérez-Gómez, 1983).

Recent research expands this understanding by demonstrating how the physical environment, when embedded with digital systems, becomes a human-scale interface that mediates interaction through sensory and behavioral feedback (Farina et al., 2018). These hybrid spaces enable architecture to respond dynamically to user activity, turning material components—such as lighting, surfaces, or furniture—into interactive systems that communicate information and emotion across physical-digital thresholds (Farina et. al., 2018). Building on this, the *Prototype Hut for the Post-Digital Age* offers a compelling demonstration of how architecture can embody both material intelligence and computational adaptability. Developed as a compact, modular dwelling, the hut integrates programmable electrochromic façades, environmental sensors, and a distributed control system that adjusts lighting, ventilation, and shading in response to real-time conditions. Rather than relying on a centralized intelligence, the design embraces decentralized responsiveness, allowing the building envelope to act as a performative, ecological interface. In doing so, it reframes architecture as an active mediator of experience—capable of balancing comfort, sustainability, and user engagement through spatial and technological synergy (Kotsopoulos et. al., 2023).

The *Connected Sustainable Home* prototype, developed at MIT, exemplifies this evolution of the built environment. Through a combination of programmable materials, embedded sensors, and model-based control systems, the architecture adapts in real-time to environmental changes and user needs. Electrochromic façades, autonomous heating and lighting regulation, and flexible modular spaces work together to turn the physical structure into an active agent of user comfort and ecological efficiency—an operational shift from static shelter to responsive ecosystem (Kotsopoulos et. al., 2023).

Emerging research also underscores the impact of responsive physical environments on human behavior and emotional well-being. Studies conducted in adaptive learning spaces and smart offices have shown that dynamic lighting, acoustic regulation, and sensor-based environmental feedback can reduce stress, support cognitive focus, and improve overall user satisfaction. These outcomes reinforce the importance of viewing the physical environment not just as a passive shell but as a behavioral interface—one capable of co-regulating with its occupants in real time. Together, these viewpoints establish a conceptual ecosystem in which the physical environment is regarded as materially anchored, sensorially rich, socially structured, and open to alteration. This relational and culturally embedded understanding of physical space provides a critical foundation for the integration of emerging technologies in urban and architectural contexts.

This relational and culturally embedded understanding of physical space provides a critical foundation for the integration of emerging technologies in urban and architectural contexts. As tools such as large language models, video recognition systems, generative AI, and immersive interfaces are deployed in the built environment, their role must be framed not as replacements or simulations of the physical, but as augmentative agents—systems that interact with, learn from, and respond to the material, sensory, and emotional qualities of space. This notion is supported by recent work by Moraitis (2024), who argues that the physical environment exists in the symbolic and perceptual shifts between architecture and landscape, the concrete and the conceptual. He emphasizes the necessity of viewing new technological integrations not just as technical enhancements, but as chances to enrich our connection to space through memory, narrative, and symbolic significance.

However, as Browning and LeCun (2022) note, the limitations of LLMs are rooted in the constraints of language itself. These systems, while powerful in generating contextually appropriate responses, operate with a shallow form of understanding that lacks the embodied, situated knowledge central to human experience. As such, integrating LLMs into spatial systems requires careful consideration of the representational gaps between linguistic models and real-world sensory perception.

In order to integrate digital systems with physical environments, it is crucial to consider it as an interpretative and relational domain. This approach facilitates the creation of adaptive, emotionally intelligent, and contextually relevant spatial systems that correspond with the intricacies of both human and environmental factors—exactly the type of spatial intelligence our research aims to foster.

*Virtual environment*

Overview of Emerging Digital Toolsets for 3D Representation and Engineering

The range of digital tools has expanded towards all megatrends within a society (Vartziotis et al., 2024). To review how immersive technologies such as Virtual Reality (VR), Augmented Reality (AR), or the combination of them – Mixed Reality (MR) are used in a physical environment, e.g. the future office, it is important to explicitly fulfill remote collaboration on physical tasks (Xiao, 2021). For example, the use of a head-worn AR display and camera can allow a remote expert to give feedback, while the integration of AR technology enables them in real-time to almost seamlessly merge and interact with virtual and physical objects. This collective operation between physical and computer-generated virtual environments is one example of many emerging solutions, which digitally increase the range of dynamics for their application (Xiao, 2021).

However, these tools often have technical limitations, such as bandwidth constraints or technological immaturity, that can impede teams' effectiveness (Kodama, 2020). One significant challenge is the absence of natural 3D representations of real humans and the difficulty in naturally manipulating virtual 3D objects (Vartziotis et al., 2024). In hybrid environments, achieving a seamless integration of physical and virtual elements can be problematic. For instance, in virtual meetings, accurately displaying participants as 3D representations that mirror their physical appearance remains a complex task. Additionally, challenges such as balancing task-specific productivity with creativity and learning, navigating uncertainty, and the lack of support in decision-making processes are still present in multiple areas of the industry (Kodama, 2020). To address the previously mentioned challenges, various exemplary solutions and toolsets have been developed across the industry (see Table 1):

Table 1: Emerging Digital Tools

| Tool Name | Description | References |
|---|---|---|
| *Apple Vision Pro* | Advanced AR headset; enhances remote collaboration through immersive experiences with lifelike 3D visualizations and interactive environments | Economist Impact (2023) |
| *Meta AI and Meta Quest 2* | Collaborative platform; facilitates real-time teamwork on intricate 3D simulations and digital twins, seamlessly integrating physical and virtual processes | Romei (2022) |
| *NVIDIA Omniverse Platform* | Collaborative platform; facilitates real-time teamwork on intricate 3D simulations and digital twins, seamlessly integrating physical and virtual processes | The future of remote work, (2023) |
| *Eve - Cloud Simulation Platform* | Cloud-based platform built on opensource components; allows for easy development of different simulation scenarios to enable iterative development and testing of products | Risi and Pronzato (2021) |
| *Tronis® & GenVis®* - | Toolset; creates an immersive adaptive virtual workspace allowing multi-user interfacing. Can be enhanced by distinct virtualized 2D interaction surfaces leveraging the co-creative potential of teams | Vartziotis et al. (2024); Phapant, et al. (2021) |
| *Collaborative Telepresence Tool* | Enables real-time communication and information exchange between remote teleworkers and physical workers; can provide near-real spatial awareness by incorporating a highly advanced stereographic camera and surround microphones, providing an accurate and detailed representation of the physical environment | Vartziotis et al. (2024); Babapour Chafi et al. (2021) |
| *CoVR Platform* | Flexible, multi-layer, client-agnostic protocol and software platform for local and remote collaborative MR. | Herrera-Pavo (2021) |

The current tools available either target a specific use or serve as aid tools to facilitate targeted tool development. However, it is necessary to intentionally direct the future innovation of workspace environment through the distinct lenses of design. Although innovation can be methodologically viewed under three scopes (feasibility, viability, and desirability) (Bultema,

2022), it is necessary to keep a holistic consideration of different scopes, which commonly affect connected megatrends within a society.

Human interaction deficits pose a significant challenge in digital workspaces. Despite the facilitation of functional communication through digital tools, these often fail to replicate the depth of human connection formed through in-person interactions. This deficit can result in adverse emotional states among remote workers, cultivating feelings of isolation and alienation, which subsequently impact workplace satisfaction and productivity (Marsh et al., 2022). The regular implementation of remote work settings diminishes spontaneous, informal encounters, such as hallway discussions and lunch breaks, which foster creativity and innovation.

To address these challenges, the development of digital workspace environments must prioritize fostering social interaction and community-building (Marsh et al., 2022). This involves creating digital platforms that encourage informal communication and the development of relationships among remote workers. To overcome current limitations, designers need to adopt holistic approaches. These should support sustainable, human-centered, and inclusive remote work cultures. Such approaches can foster creative solutions for future challenges and benefit all levels of industry—from large corporations to small and medium-sized enterprises (Kodama, 2020). However, the commonly used tools for collaboration among distributed teams are still heavily reliant on 2D computer screens for interaction and information presentation. While this method is suitable for decision-making, it presents a significant obstacle for co-creative collaboration, idea generation, and design. These activities are inherently different when experienced in 2D and require the conveyance of proper spatial and social cues, as well as rich channels of perception.

While digital workspaces offer numerous advantages, addressing human interaction deficits is critical for maintaining employee well-being and fostering creativity and innovation. There are six individual factors, which serve as moderators for negative effects of interaction deficit: personality, age, gender, work and technology presences, computer self-efficiency, mindfulness (Marsh et al., 2022). Employers should seek to take into account these factors in any design of new solutions to guarantee noble results. By developing more immersive digital tools and adopting organizational practices that emphasize regular, meaningful communication, remote work environments can become more effective and supportive of their workforce.

*Fusion of Environments*

As remote work reshapes the boundaries between domestic and professional life, the architectural challenge is no longer about designing either physical or digital spaces—but rather about orchestrating transitions, overlaps and hybrid rhythms across both. The fusion of environments refers to the emergence of continuous, dynamic conditions where meaning, attention and effect move fluidly between embodied space and virtual interface. Large Language

Models (LLMs), in this context, do not merely automate tasks or respond to commands but instead operate as real-time semantic mediators. Through their ability to parse natural language, infer context and generate intent, LLMs become spatial interlocutors— interpreting the user's emotional and cognitive state and proposing micro-interventions in both physical and virtual space. Rather than replacing architectural form in either physical or virtual environments, LLMs suggest spatial possibilities: repositioning objects to support rhythm, modulating light to reflect internal tempo or adjusting digital environments based on subtle behavioral cues. They enable a reframing of the built environment not as static infrastructure but as a responsive interface that can mirror and modulate the user's experience. In this fusion, architecture becomes a conversation and space becomes a co-agent in the shaping of attention, mood and presence. This section sets the ground for understanding how adaptive environments might be designed not through pre-defined logic but through open-ended interactions with language-based AI— enabling a co-constructed experience of space across domains.

Defining Fusion

Following the global transition to remote and hybrid work, residential environments have transformed into venues of professional engagement. We operate from our kitchens, participate in meetings from our bedrooms, and manage deadlines concurrently with daily life. This convergence disrupts traditional divisions between physical and digital environments, affecting both function and our perception, design, and occupation of space (Felstead and Reuschke, 2020). Historically, architecture has concentrated on either the physical domain or the digital interface: edifices vs platforms, interiors versus displays. At present, these categories collapse. The home office is not merely a room with a laptop; it is a hybrid condition where presence, cognition, and interaction simultaneously navigate both physical and digital realms (Waizenegger et al., 2020).

Commonly termed "phygital" workspaces, these settings enable the fluid integration of physical and digital components to improve productivity, collaboration, and well-being (Lo et al., 2024). Yet, the fluidity they enable can also strain cognitive and emotional boundaries. Remote workers sometimes shift their focus between a bright screen and a distraught child, as well as between scheduled meetings and domestic interruptions (Waizenegger et al., 2020). During the pandemic, homes were no longer just places of shelter but infrastructural sites of life, work and learning, occupied "24 hours a day" (Babapour Chafi et al., 2022). This environmental fusion matters because most spaces were not designed for this entanglement. While traditional offices segregated domains—work, home, leisure—the contemporary hybrid condition blends them. And yet, most architectural systems remain static, unable to interpret or adapt to this emerging complexity.

To move beyond this disconnect, design must treat physical space and digital interface as a unified ecology. Fusion demands architectural responses that are attuned to flow, co-presence and transition rather than fixed use (McCullough, 2005), (de Souza e Silva, et al. 2025). Therefore, we define environmental fusion as the entangled coexistence and cofunctioning of physical and digital environments into a single, dynamically adaptive field. Unlike layering—where digital tools sit atop inert structures—fusion implies mutual transformation: the digital reshapes how the physical is perceived and used while the physical grants friction, context and atmosphere to the digital (Cache, 1995; Kitchin and Dodge, 2011)

Fusion, then, is not a theoretical abstraction. It is the lived spatial condition of millions of remote and hybrid workers. As such, design must rise to meet it—not as a split between physical and digital but as a shared experience to be carefully mediated, supported and shaped (Xu et al., 2024).

**Generative AI Tools for Digital and Virtual Environments**

Generative AI is rapidly reshaping digital and virtual environments by automating content creation, enhancing immersion, and enabling intelligent interactivity. Modern frameworks leverage large language models (LLMs), diffusion networks, and GANs to generate 3D assets, textures, and entire scenes directly from natural language or image prompts, substantially reducing manual modeling efforts and accelerating design iterations [41]. Systems like VRCoPilot demonstrate how users can sketch or speak layout ideas within a VR space, with the AI dynamically completing and refining 3D layouts in real time, fostering a mixed-initiative design process that blends human creativity with AI assistance (Chen et al., 2023). In mixed reality (MR), models like sMoRe interpret spoken or typed instructions to generate, place, and manage virtual objects tailored to physical surroundings, bridging cognitive intention and spatial action (Singhal, 2022). Furthermore, innovations like SpaceBlender integrate live 2D environments into contextually enriched 3D telepresence settings, facilitating more immersive and collaborative virtual places that mirror a user's actual surroundings (Guo et. al., 2023). Early studies in personalized AI narration in virtual reality—where adaptive, GPT-driven voice agents modify information according to user interaction—have demonstrated significant enhancements in engagement and sustained attention during cultural learning tasks (Kim et al., 2022). Taken together, these tools indicate a paradigm shift: from static virtual scenes to dynamic, responsive environments co-authored by users and generative AI—unlocking new possibilities for design, collaboration, and education in digital workspaces.

*The Role of LLMs in Mediating Space Through Language*

As environments become increasingly hybrid, they demand more than automated triggers or smart interfaces. They require systems capable of interpreting human expression, intention and affect. LLMs are deep neural networks trained on massive corpora of human generated text to predict and generate language in a way that simulates human communication. At their core, they operate through next-token prediction: given a sequence of tokens, the model computes the probability of the next word. However, the Transformer architecture (Vaswani et al., 2017) allows these models to attend across long contexts and infer semantic, logical and even affective relationships. This makes them capable of dialogic reasoning, context maintenance and responsive adaptation to subtle user cues. Although they do not "understand" in a human sense, their ability to generate contextually fluid, semantically rich responses renders them uniquely capable of translating fuzzy or emotional language into spatial implications (Bommasani et al., 2021). Rather than executing direct commands, LLMs offer semantic mediation—interpreting vague, affective or ambient expressions and translating them into spatial adjustments. For instance, "I feel overwhelmed" might trigger a softening of lighting or reduction of acoustic stimuli.

LLMs support this mediation through several key affordances. First, through affective inference, LLMs can detect user states from indirect language and respond with spatial interventions. Systems like LaMDA, for instance, demonstrate the ability to infer mood and adjust environments accordingly (Thoppilan et al., 2022; Li et al., 2023).

Second, they enable narrative translation to spatial strategies, serving as a bridge between abstract user goals—such as "I want to feel more focused"—and concrete spatial interventions like acoustic zoning or light modulation. Finally, through multimodal understanding, LLMs, when linked with visual, gestural, or sensor inputs, can guide real-time adaptations across both physical and XR environments (Xu et al., 2024; Alayrac et al., 2022).

As these systems are increasingly embedded into spatial interfaces, from smart homes to virtual collaboration environments, they shift our interaction paradigm: from tool-based command systems to ambient, co-creative mediation layers. In this context, the role of the designer is not to predefine every interaction but to choreograph potential exchanges between language, space and responsiveness.

*Designing Co-Adaptive Environments: From Interface to Atmosphere*

If LLMs provide a new form of semantic mediation between user and space, what design paradigm emerges from this capacity? We argue that LLMs enable a shift from space as interface —a controllable surface—to space as atmosphere—a co-adaptive, expressive field. Unlike traditional smart systems that respond to direct commands or sensor inputs, LLMs invite conversational engagement. A phrase like "This space feels too chaotic today" is not a directive but an affective expression. Yet an LLM can interpret it as an environmental cue—adjusting lighting, reducing distractions or offering calming visual overlays.

This transition reframes architectural design. No longer limited to defining forms or automations, the designer orchestrates a system of relationships—between language, mood and environment. The space does not merely "function." It listens, interprets and participates in the ongoing shaping of user experience. LLMs support this transformation through three interconnected capabilities. First, co-creative intelligence allows users in hybrid and remote settings to engage in real-time spatial generation and adjustment via dialogue. Tools like VRCoPilot and SpaceBlender facilitate this process, enabling users to co-author their environments through natural language interactions and shifting the role of the interface from passive input to active collaboration (Chaudhuri et al., 2023; Wang et al., 2023).

Second, empathic adaptation enables LLM-mediated environments to interpret emotional tone and cognitive states, adjusting spatial elements to enhance focus, reduce stress, or accommodate neurodiverse sensitivities. Recent studies emphasize how such adaptive systems can contribute significantly to well-being in workspaces (Woodbury et al., 2024). Finally, atmospheric sensing is achieved through multimodal LLMs such as Video-ChatGPT and LongVLM, which integrate language with video, sound, and gesture inputs. These systems make it possible to sense the affective atmosphere of an environment, enabling responsive spatial configurations that evolve not only with what is said, but also with how it is expressed (Zhang et al., 2024) – (Jung et al., 2023)

Such capabilities position the LLM not as a background engine but as a participant in space-making. Design shifts from scripting functions to cultivating relationships between presence, language and adaptation. For remote and hybrid work, this suggests a profound spatial reconnection—one where architectural systems mirror and respond to inner states, fostering environments that grow with us.

It should be noted that such systems despite their great potential also raise some important concerns. When machines analyze emotions, there is a possibility that they will misread subtleties, reinforce bias, or violate individuals' right to privacy. The European Union's proposed Artificial Intelligence Act warns against the unregulated use of emotion recognition in sensitive situations such as education or employment [European Commission, "Proposal for a Regulation laying down harmonised rules on Artificial Intelligence (Artificial Intelligence Act)," 2021]. A transparent data governance system, user permission, and equitable design frameworks are essential components in order to guarantee that AI-enhanced environments continue to be courteous and welcoming to all individuals.

In what follows, we explore how such a paradigm might take form in practice. To ground the above discussion in a practical scenario, we undertook a prototypical experiment—an Adaptive Workpod—as a testbed for co-adaptive spatial fusion.

**Case Study: An Adaptive Workpod Prototype**

We defined an adaptive workpod as a flexible, self-contained workspace unit designed to dynamically respond to user needs, environmental conditions, or task-specific requirements. We evaluated the adaptive workpod in four remote work sessions per participant. During each session, the system continuously logged three streams of data: user utterances captured via an open microphone, LLM inferences produced by ChatGPT 4o in advanced voice-and-vision mode, and the resulting adaptations enacted by actuators (devices that convert signals into motion). Multimodal inputs consisted of 1) audio—for natural verbal expressions such as "I feel overwhelmed" or "I need to focus"; 2) video—for engagement indicators like gaze direction, posture shifts, or facial fatigue and 3) optional browser and application log activity to infer focus drift or variations in work intensity. The LLM agent processed these cues in real time, drawing on its short-term memory of prior interactions to generate semantic recommendations. Those recommendations drove three types of actuators: ambient lighting adjustments in brightness and color temperature, screen display modes (for example, full-screen immersion versus a low-stimulation interface), and sound-masking output such as white noise or ambient sounds. After each session, participants rated every adaptation as helpful, intrusive, or irrelevant. We then computed three evaluation metrics: self-reported focus and stress levels measured immediately before and after each adaptation; perceived appropriateness of each adaptation based on user ratings; and personalization over time, reflecting whether the system's memory produced increasingly tailored responses across sessions.

Designed for remote work contexts, the Adaptive Workpod integrates a Large Language Model (LLM) with simple multimodal inputs (microphone, webcam, browser activity) and smart outputs (ambient light, screen mode, sound masking) to create a responsive home workspace that adapts to the user's emotional and cognitive state in real time. The objective is to investigate whether LLM-mediated semantic cues and video-derived engagement signals can trigger spatial adaptations that improve focus, reduce stress, and increase user well-being during remote work. For the experiment, we used ChatGPT 4o, with advanced voice-and-vision mode, a natural and emotional sounding voice assistant with vision capabilities. The inputs included three primary channels. First, *audio* was captured via an open microphone to allow for natural verbal expressions (e.g., spontaneous utterances such as "I feel overwhelmed" or "I need to focus"), enabling the LLM to detect and interpret emotional states through language. Second, *visual cues* of engagement and exhaustion, such as gaze direction, posture changes, and face microexpressions, were obtained via webcam video input. We employed the publicly accessible ChatGPT (GPT-4o) model to instantiate the LLM agent through the Application Programming Interface (API). It processed verbal and behavioral cues to infer user intent and maintained short-term memory of prior interactions to provide contextually relevant and personalized responses.

While local deployment offers advantages in privacy and latency, our prototype relied on cloud-based access to leverage the advanced voice and vision capabilities of GPT-4o, which were essential for real-time semantic interpretation and multimodal input fusion.

*Experimental inputs*

- Audio: open microphone for natural verbal expressions (e.g., "I feel overwhelmed", "I need to focus").
- Video: webcam-based engagement indicators such as gaze direction, posture shifts, or facial fatigue.
- Optional browser/log activity to infer focus drift or work intensity.

*LLM Tool*

A lightweight, locally hosted or API-based LLM model (e.g., GPT-4o) that processes verbal and behavioral cues and infers intent using memory of past interactions.

*Actuators*

- Smart light bulb: adjustable color temperature and brightness.
- Sound system: background music on/off, white noise, or silence.
- Desktop interface: distraction-reducing overlays or focused UI states.

*Workflows and Outcomes*

1.

We discuss below four different participant (worker) behaviors (cues); how they were interpreted and responded to by the LLM tool; and how the interventions were evalauted by the researchers.

1. <u>Drowsiness Recovery</u>

User cue: "I'm feeling a bit drowsy."

LLM interpretation: detects lowered energy and risk of mind-wandering.

Actuator action: a recommendation to switch ambient lights to a cooler, brighter white and suggestion of a two-minute standing stretch are displayed via an on-screen prompt.

Measured outcomes: a one-level increase in self-reported alertness on a five-point scale, participant's posture returning to upright within thirty seconds, and participant's rating the intervention as helpful.

2. Focus Restoration

User cue: eyes gaze away from the screen for more than ten seconds.

LLM interpretation: infers loss of visual focus or fatigue.

Actuator action: a brief on-screen reminder ("Take a 30-second break") is shown, and a recommendation to dim peripheral lighting to reduce glare.

Measured outcomes: 50% reduction in off-screen gaze events over the next five minutes, system latency under one second, and intrusive ratings below 10%.

3. Distraction Mitigation

User cue: two consecutive five-minute visits to social media sites.

LLM interpretation: detects procrastination or distraction.

Actuator action: recommendation to block social media for five minutes and play soft white noise. Measured outcomes: return to work-related domains within two minutes, at least a one-point increase in self-reported focus, and 80% of interventions rated as helpful by particpant.

4. Stress Alleviation

User cue: "This task is stressing me out."

LLM interpretation: high stress signal requiring calming intervention.

Actuator action: recommendation to gradually lower the light to a warmer hue and to follow a guided breathing exercise via the voice assistant.

Measured outcomes: at least a one-point drop in self-reported stress within three minutes, completion of the breathing prompt, and recall of this intervention within ten seconds of similar cues in later sessions.

**Conclusion**

This paper explored how large language models can mediate adaptive architectural responses in the context of individual work environments. We introduced a lightweight prototype—the Adaptive Workpod—that processes multimodal cues such as speech, gaze, and behavioral patterns to generate context-aware recommendations via a publicly available LLM. Our methodology outlined a non-invasive evaluation framework, emphasizing focus, stress, appropriateness, and personalization without requiring biosensing or deep instrumentation. Through illustrative workflows, we demonstrated how the system interprets affective states and offers targeted interventions across common cognitive and emotional challenges.

These findings support the idea that meaningful affective adaptation can be achieved through dialogical interaction and naturalistic cues, paving the way for a new class of emotionally attuned workspaces. Future developments will require enabling actuation, expanding user modeling, and integrating physiological or collaborative sensing. As adaptive environments become more intelligent, maintaining user agency, privacy, and transparency will be critical. We argue that this form of computational mediation—grounded in lived language—extends the boundaries of architectural responsiveness and offers a promising trajectory for the design of emotionally resonant digital-physical workspaces.